%% file: main.tex
\documentclass[sigconf,final,screen]{acmart}
\usepackage[utf8]{inputenc}
\usepackage[T1]{fontenc}
\usepackage{amsmath,amsfonts,nicefrac}
\usepackage{hyperref,url,booktabs,multirow,microtype}
\usepackage{enumitem,algorithm,algorithmicx,algpseudocode}
\usepackage{xcolor,graphicx}

\newcommand{\mr}[2]{\multirow{#1}{*}{#2}}
\newcommand{\mc}[3]{\multicolumn{#1}{#2}{#3}}

\algblockdefx{WhileAsync}{EndWhileAsync}
    [1]{\textbf{async while} #1}
    [0]{\textbf{end while}}
\setlist[itemize]{leftmargin=12pt,topsep=2pt}
\setlist[enumerate]{leftmargin=12pt,topsep=2pt}

\newcommand{\oducs}{\department{Department of Computer Science}\institution{Old Dominion University}\city{Norfolk}\state{VA}\country{USA}}

\title{StreamingHub: Interactive Stream Analysis Workflows}

\author{Yasith Jayawardana}
\orcid{0000-0001-5992-6818}
\affiliation{\oducs}
\email{yasith@cs.odu.edu}

\author{Vikas G. Ashok}
\orcid{0000-0002-4772-1265}
\affiliation{\oducs}
\email{vganjigu@odu.edu}

\author{Sampath Jayarathna}
\orcid{0000-0002-4879-7309}
\affiliation{\oducs}
\email{sampath@cs.odu.edu}

\begin{abstract}
\input{sections/0.abstract}
\end{abstract}



\copyrightyear{2022}
\acmYear{2022}
\setcopyright{rightsretained}
\acmConference[JCDL '22]{The ACM/IEEE Joint Conference on Digital
Libraries in 2022}{June 20--24, 2022}{Cologne, Germany}
\acmBooktitle{The ACM/IEEE Joint Conference on Digital Libraries in 2022
(JCDL '22), June 20--24, 2022, Cologne,
Germany}
\acmDOI{10.1145/3529372.3530936}
\acmISBN{978-1-4503-9345-4/22/06}

\begin{document}

\maketitle

\section{Introduction}
\input{sections/1.introduction.tex}

\section{Related Work}
\input{sections/2.relatedwork.tex}

\section{System Design}
\input{sections/3.design.tex}

\section{Evaluation and Results}
\input{sections/4.experiments.tex}

\input{sections/5.evaluation.tex}

\section{Discussion}
\input{sections/6.discussion.tex}

\section{Conclusion}
\input{sections/7.conclusion.tex}

\section*{Acknowledgments}
\noindent This work was supported in part by NSF CAREER-245523.
Any opinions, findings and conclusions or recommendations expressed in this material are the author(s) and do not necessarily reflect those of the sponsors.

\balance
\bibliographystyle{ACM-Reference-Format}
\bibliography{references}

\end{document}

%% file: sections/0.abstract.tex
Reusable data/code and reproducible analyses are foundational to quality research.
This aspect, however, is often overlooked when designing interactive stream analysis workflows for time-series data (e.g., eye-tracking data).
A mechanism to transmit informative metadata alongside data may allow such workflows to intelligently consume data, propagate metadata to downstream tasks, and thereby auto-generate reusable, reproducible analytic outputs with zero supervision.
Moreover, a visual programming interface to design, develop, and execute such workflows may allow rapid prototyping for interdisciplinary research.
Capitalizing on these ideas, we propose \textit{StreamingHub}, a framework to build metadata propagating, interactive stream analysis workflows using visual programming.
We conduct two case studies to evaluate the generalizability of our framework.
Simultaneously, we use two heuristics to evaluate their computational fluidity and data growth.
Results show that our framework generalizes to multiple tasks with a minimal performance overhead.

%% file: sections/1.introduction.tex
Scientific research is cumulative in nature; discoveries are made upon prior findings to broaden our understanding of the world.
Often, this involves iteratively collecting data and developing, validating, and executing multiple analyses leading to tangible results.
With many researchers exploring similar ideas, the landscape of scientific research is highly competitive \cite{fox1992research}.
As a result, researchers tend to overlook the aspects of data reusability, experiment reproducibility, and result verifiability in the interest of time.
To exemplify, studies have shown that most scientific publications fail to report certain information needed to replicate experiments \cite{nekrutenko2012next} and reproduce analyses \cite{ioannidis2009repeatability}, leading to an increasing number of retracted papers \cite{steen2011retractions} and failing clinical trials \cite{prinz2011believe}.
Naturally, this has led to discussions on how researchers, institutions, funding bodies, and journals can establish guidelines to increase the reusability and verifiability of research assets \cite{sandve2013ten}.
As a result, principles such as FAIR \cite{wilkinson2016fair} have emerged to guide researchers toward making assets \textbf{F}indable, \textbf{A}ccessible, \textbf{I}nteroperable, and \textbf{R}eusable through the inclusion of informative metadata.
While this works in theory, manually generating FAIR metadata is a time-consuming task \cite{peng2011reproducible}, meaning researchers are prone to overlooking it when FAIRness is optional.
\textbf{Hence, automating the process of propagating reusable metadata to facilitate reproducibility, has practical value.}

Another time-consuming issue in research is having to ``reinvent the wheel'' when existing code is not reusable.
This problem can partly be avoided by writing modular code that generates deterministic results.
However, as the research gets more complex, so does managing the code base.
Alternatively, one could resort to using \textit{Scientific Workflow} (SWF) systems \cite{goble2020fair}, which adapt the flow-based programming paradigm \cite{morrison2010flow} to model complex data transformations as a Directed Acyclic Graph (DAG) of simple, reusable data transformations (i.e., a workflow).
SWF systems allow researchers to build and manage complex workflows, and run parameter-driven simulations and alternative experimental setups with ease.
Among the multitude of SWF systems \cite{goble2020fair,liew2016scientific}, some are code-based (e.g., Pegasus), while others are based on \textit{visual programming} \cite{burnett1995visual} (e.g., Kepler, KNIME, Node-RED).
Code-based SWFs are challenging for scientists with no programming background to use \cite{brimhall2001removing}, whereas SWF systems based on visual programming abstract away the complexity of underlying programming with a conceptual, visual design.
However, they provide no mechanism to intelligently consume metadata and propagate metadata across data transformations by default.
\textbf{Hence, integrating metadata propagation into SWF systems would allow researchers to build reusable workflows that transform data into reusable, reproducible analytics.}

In this paper, we propose \textit{StreamingHub}, a framework to build metadata propagating, interactive stream analysis workflows using visual programming.
In the interest of accessiblity, we implement this framework on Node-RED, and use Data Description System (DDS) as our baseline metadata format (see Section~\ref{sec:dds}).
StreamingHub allows to propagate metadata in three forms: (a)~from data sources into workflows, (b)~between data transformations in a workflow, and (c)~from workflows into analytic outputs.
Within workflows, it records provenance information with zero supervision.
We demonstrate how StreamingHub combines the ideas of reusable data, reusable workflows, and metadata propagation to auto-generate reusable, reproducible analytic outputs and thereby simplify research.
Our contribution is three-fold:
\begin{enumerate}
\item We propose an extensible metadata format (DDS) to describe \textit{data sources}, \textit{data sets}, and \textit{data analytics}.
\item We demonstrate how we built interactive workflows on StreamingHub for two distinct stream analysis tasks.
\item We evaluate the performance of workflows generated from StreamingHub, and discuss our findings.
\end{enumerate}

%% file: sections/2.relatedwork.tex
In this section, we first document current metadata standards for file-based and stream-based data, and several data sharing platforms that index them.
Second, we document several systems for stream processing, and how they were evaluated.
Third, we document several SWF systems and workflow engines that run SWFs at scale.

\subsection{Metadata Standards}

Metadata standards such as MARC \cite{avram1968marc} by the Library of Congress, Dublin Core \cite{kunze2007dublin} by the Dublin Core Metadata Initiative (DCMI), and PMH and ORE  by the Open Archive Initiative (OAI) \cite{lagoze2008object}, provide domain-agnostic metadata elements for describing digital assets.
However, domain-specific metadata standards such as HCLS \cite{dumontier2016health} for healthcare and life science data, ESML and EML \cite{fegraus2005maximizing} for ecological data, GeoSciML \cite{sen2005geosciml} for geological data, and METS \cite{cundiff2004introduction} for digital library data, also exist.
The key difference between domain-agnostic and domain-specific metadata standards is in their level of specificity; the former provides generic attributes, while the latter provides attributes specific to a particular domain.
Regardless, these metadata standards are based on markup languages such as XML, JSON, YAML, or RDF.

Platforms such as FAIRsharing~\cite{sansone2019fairsharing}, FigShare~\cite{thelwall2016figshare}, Dataverse~\cite{king2007introduction}, and DataHub~\cite{bhardwaj2015datahub} facilitate the public distribution of open research data, and the selective distribution of confidential/copyrighted data for educational and collaborative purposes.
FAIRsharing, in particular, indexes over 1546 metadata standards, 1800 databases, and 146 policies from the natural science, engineering, humanities and social science domains \cite{fairsharingstds}.
Moreover, spatio-temporal data storage platforms such as Galileo \cite{malensek2011galileo} mandates the use of certain attributes in metadata for efficient indexing and query evaluation.
This includes domain-specific spatial metadata such as geo-coordinates, temporal metadata such as timestamps, identity metadata such as device identifiers, error-detection metadata such as checksum, and user-defined metadata such as task-specific attributes.
However, a 2015 study \cite{roche2015public} reports that 56\% of natural sciences datasets had incomplete metadata, and 64\% of them were archived in a manner that hindered data reuse.
This suggests that while a plethora of metadata standards exist, a large proportion of digital assets lack metadata that facilitates reuse.
It is thus imperative to establish routines that evaluate the quality and completeness of metadata.
The FAIR Principle~\cite{wilkinson2016fair}, for instance, is a well-established guideline to quality-check metadata based on findability, accessibility, interoperability, and reusability.
Therefore, to describe data, one should preferably use FAIR, domain-agnostic metadata standards that are extensible for domain-specific use cases.
\textit{In this work, we use the Data Description System (DDS) \cite{jayawardana2019dfs,jayawardana2020streaming} metadata format to describe data sets, data streams, and data analytics.}

\subsection{Metadata for Streaming}
While data analysis was originally performed offline as batch jobs, time-sensitive applications such as stock prediction demand a shift towards stream-processing and online analysis.
Frameworks such as \textit{Apache Flink}, \textit{Apache Beam}, and \textit{Google DataFlow} support building both batch processing and stream processing workflows \cite{akidau2015dataflow}.
These frameworks, while particularly being geared towards stream-processing, support batch-processing as a special case of stream-processing.
Studies show that having stream-level metadata could improve analysis performance \cite{alvarez2021generic,le2005estimating}.
As such, these frameworks provide APIs to access and manage stream-level metadata.
Moreover, libraries such as \textit{LabStreamingLayer} \cite{kothe2014lab} and \textit{IFoT} \cite{yasumoto2016survey} facilitate the transmission of metadata alongside data streams.
LabStreamingLayer, for instance, is an embeddable library for network discovery, time-synchronization, and low-latency streaming, whose APIs are already used by several device manufacturers to stream sensory data.
LabStreamingLayer uses an XML-based metadata format named XDF which provides schema for generic audio/video data, and for sensory data including EEG and Gaze.
Similarly, IFoT uses a metadata format with attributes such as data type, granularity, location, and tags, to unify the processing of information flows.
Both libraries handle data and metadata independently for efficient query evaluation and selective data streaming.
\textit{In this work, we use LabStreamingLayer to stream live data, replayed data, and synthetic data along with DDS metadata into stream analysis workflows.}

\vspace{1em}
\noindent\textbf{Evaluating Streaming Performance:}
Latency, throughput, scalability, and resource utilization, specifically of memory and CPU, are commonly used measures of stream processing performance.
One study \cite{van2020evaluation} used them to compare the performance of stream processing frameworks under different data volumes, workloads (constant rate/bursty) and workload complexities (low/high).
Another study \cite{nasiri2019evaluation} used latency, throughput, scalability, and resource utilization to compare distributed stream processing frameworks under different data volumes, using applications from Yahoo Streaming Benchmarks \cite{chintapalli2016benchmarking} and RIoTBench \cite{shukla2017riotbench}.
\textit{In this work, we use latency to analyze the stream analysis performance under different data volumes, workloads, and workload complexities.
We also propose two heuristics aimed at identifying computational bottlenecks in a transformation (fluidity), and measuring the change in data volume through a transformation (growth factor).
}

\subsection{Scientific Workflow Systems}
Mature SWF systems, such as Pegasus \cite{deelman2015pegasus} and Kepler \cite{altintas2004kepler}, are widely used among scientific communities.
In Pegasus, workflows are described using the Directed Acyclic Graph XML (DAX) format. It also provides APIs (Python/Java/R) to define workflows in terms of \textit{replicas} (where data is located), \textit{sites} (where workflows are deployed), and \textit{transformations} (what to do with data). A transformation specifies which files to execute and how to invoke them, to complete a portion of the workflow.
Workflows defined using these APIs are later compiled into DAX format.
Kepler workflows, on the other hand, are designed through a visual-programming interface, where scientists sketch each step of the workflow (i.e., \textit{actor}) and wire them together.
These workflows are described using the Modeling Markup Language (MoML) \cite{lee2000moml}.
Common Workflow Language (CWL) \cite{amstutz2016cwl} is a SWF notation for describing SWFs regardless of their runtime.
SWF runtimes such as Apache AirFlow, Apache Taverna, and Apache Airavata provide support for CWL. Pegasus also provides a converter to transform CWL workflows into Pegasus YAML format. Both Pegasus and Kepler traditionally support batch processing. However, Kepler has been used for streaming data collections as well. Both systems are file-based, which poses architectural limitations on adapting them for real-time streaming applications.

When executing SWFs at scale, scientists often require computational resources beyond what is available on a single computer.
Traditionally, SWFs were executed at scale on institutional HPC clusters (or grids).
Nowadays, scientists have access to multiple infrastructures such as grids, public clouds, and private clouds to execute their SWFs \cite{foster2008cloud}.
In particular, Grid Computing infrastructures such as Open Science Grid \cite{pordes2007open}, TeraGrid \cite{catlett2008teragrid}, and XSEDE \cite{towns2014xsede} provide computational resources for scientists to run workflows.
Public clouds such as AWS, and private clouds such as FutureGrid \cite{von2010design} have also been used to execute workflows.
Pegasus workflows can be run at scale on any environment running HTCondor \cite{thain2005distributed}, and more recently, on AWS Batch \cite{awsbatch}.

\textit{Orange} \cite{demvsar2013orange}, KNIME \cite{berthold2009knime}, VisTrails\footnote{This project is no longer maintained}\cite{bavoil2005vistrails}, \textit{NeuroPype} \cite{neuropype}, \textit{Neuromore} \cite{neuromore}, \textit{Porcupine} \cite{van2018porcupine} and \textit{Node-RED} \cite{nodered} also provide visual-programming interfaces to design SWFs.
Orange, KNIME, and VisTrails are geared towards exploratory data analysis and interactive data visualization, while NeuroPype, Neuromore, and Porcupine are geared towards neuroimaging applications. Orange is a standalone application, and does not support distributed execution of workflows.
NeuroPype workflows can be run at scale on \textit{NeuroScale} \cite{rosenboom2019more}.
Similarly, workflows created in Neuromore and Porcupine can be run at scale on \textit{NiPype} \cite{gorgolewski2011nipype}, a neuroimaging data processing framework which supports distributed execution of workflows.
In terms of data, Orange, KNIME, and VisTrails lack support for online stream-processing, while NeuroPype, Neuromore, and Porcupine support it.
Yet, they are specialized for neuroimaging applications, and do not generalize well to other domains.
Node-RED, on the other hand, is more generic, and allows to build event-driven applications and deploy them locally, on the cloud, and on the IoT \cite{nodered}.
Workflows created in Node-RED are portable, and can be exported (in JSON format), imported, and deployed among instances \cite{nodered}.
\textit{In this work, we use Node-RED as the visual programming front-end to build data analysis workflows and to import/export workflows in JSON format.}

%% file: sections/3.design.tex
\begin{figure*}[t]
\centering
\fbox{
\includegraphics[width=.95\linewidth]{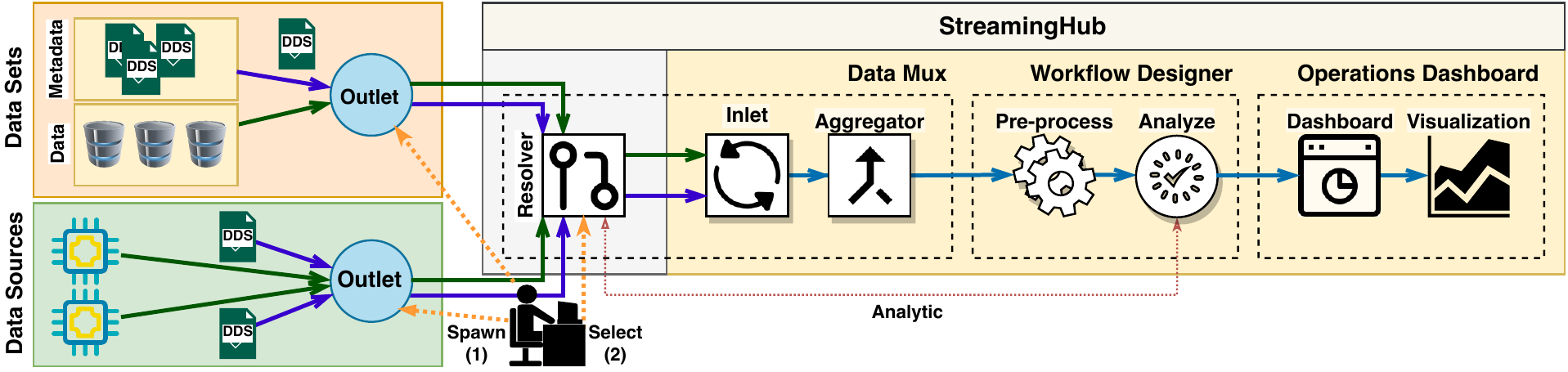}
}
\caption{
\textbf{Component-level Overview of StreamingHub:}
\textnormal{The \textit{DataMux} uses \textit{DDS} metadata to convert data from connected sensors and datasets into data-streams.
This data is then passed into the \textit{Workflow Designer} to facilitate the design and execution of workflows.
The results generated by running workflows are displayed in real-time on the \textit{Operations Dashboard}.
}
\label{fig:architecture}}
\end{figure*}

\subsection{Data Description System (DDS)}
\label{sec:dds}
DDS is a collection of schemas to create metadata for describing data-sources, data-analytics, and data-sets.
\textbf{Data-source} metadata provides attributes to describe data streams, such as their frequency and channels.
\textbf{Data-analytic} metadata provides attributes to describe analytic data streams and their provenance (i.e., the hierarchy of transformations that lead to it).
\textbf{Data-set} metadata provides attributes to describe the ownership, identification, provenance, and groups (i.e., viewpoints) of a dataset, and to reference a \textit{resolver} script that maps stream-wise and/or group-wise queries into data. Table~\ref{tab:DDS-schemas} provides a summary of the metadata attributes used in DDS.

\begin{table}[ht]
\centering
\caption{Summary of metadata attributes used in DDS\label{tab:DDS-schemas}}
\small
\input{tables/DDS.attrs.tex}
\end{table}

\begin{figure*}[t]
\centering
\fbox{
\includegraphics[width=.46\linewidth]{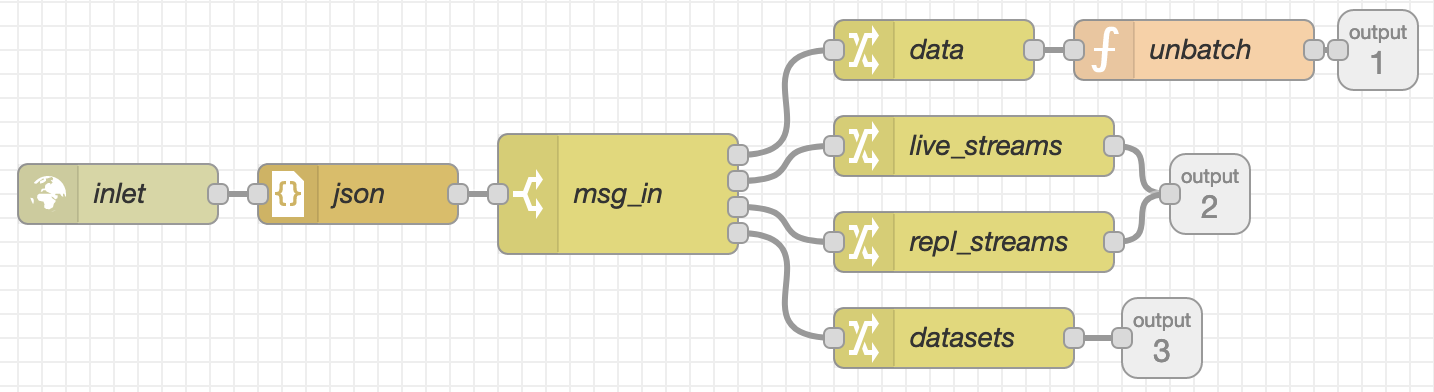}
}
\fbox{
\includegraphics[width=.46\linewidth]{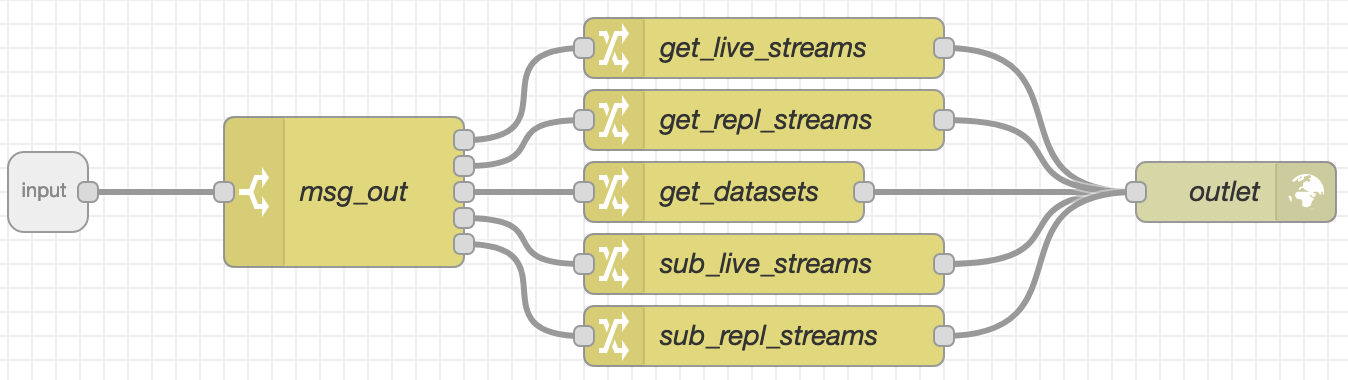}
}
\caption{
\textbf{DataMux API Implemented on Node-RED:}
\textnormal{The \textit{DataMux In} (left) sub-flow reads incoming messages from DataMux and routes them into the correct path in the workflow.
The \textit{DataMux Out} (right) sub-flow parses messages generated from the workflow and converts them into the format expected by DataMux.}
\label{fig:subflow}
}
\end{figure*}

\subsection{DataMux}
DataMux (see Figure~\ref{fig:architecture}) operates as a bridge between connected sensors, datasets, and data-streams.
It uses DDS metadata to create the data-streams needed for a task, and supports three modes of execution:
\begin{itemize}
\item\textbf{Live Mode} -- generate a data-stream that proxies live data from connected sensors (using \textit{data-source} metadata)
\item\textbf{Replay Mode} -- generate a data-stream that replays a stored dataset (using \textit{data-source} and \textit{data-set} metadata)
\item\textbf{Simulate Mode} -- generate a data-stream of guided/unguided synthetic data (using \textit{data-source} metadata)
\end{itemize}
In Live Mode, the DataMux uses LabStreamingLayer\footnote{\url{https://github.com/sccn/labstreaminglayer}} \cite{kothe2014lab} to interface with live data.
The underlying functionality of each mode is encapsulated within a WebSocket API, i.e., the \textit{DataMux~API}.
It serves as an interface to stream data into workflows and back.\\

\noindent\textbf{Usage:}
The user first spawns the DataMux (server) on the local network, and opens a WebSocket connection to it.
This DataMux connection is initially in \textit{awaiting} state.
Using this connection, the user can query for available live data-streams (which are generated from connected sensors in the local network) and replay data-streams (which are generated from datasets on the local file system).
When queried for live data-streams, the DataMux discovers LabStreamingLayer \textit{outlets} on the local network, and returns a list of them.
When queried for replay data-streams, the DataMux first discovers DDS data-set metadata files on the local file system, and returns a list of their data-streams by calling the \textit{resolver} script that each metadata file points to.
From these returned data-streams, the user can subscribe to all or a subset of them, and start receiving their data.
Note that data-stream discovery and subscription are kept independent to improve scalability.
When a user subscribes to a data-stream, the DataMux connection transitions into \textit{streaming} state, and spawns \textit{inlets} to receive data.
For live data-streams, each \textit{inlet} connects to a LabStreamingLayer \textit{outlet}, and proxies data from it in real-time.
For replay data-streams, however, each \textit{inlet} connects to one or more files in a dataset, and replays its data at the frequency it was recorded in.
Next, data from these inlets pass through an \textit{aggregator} which performs time-synchronization and merges data-streams where applicable.
When merging data-streams, the \textit{aggregator} adheres to the data frequencies specified in the metadata, to facilitate temporally-consistent replay and simulation.
When the DataMux connection is in \textit{streaming} state, the subscribed data-streams will continue streaming until the connection ends, or until the data-streams end.

\subsection{Workflow Designer}
The workflow designer is the visual programming front-end to build scientific workflows (see Figure~\ref{fig:ema-workflow} for an example).
We use Node-RED to implement the workflow designer, as it offers a visual programming front-end to build workflows, while allowing users to import/export workflows in JSON format.
This facilitates both technical and non-technical users to design workflows and share them among peers.
We implemented a set of nodes to interface with the \textit{DataMux} and made them available on the workflow designer.
Workflows generated using this interface comprise of \textit{transformation nodes} and \textit{visualization nodes} that are bound in a directed graph using \textit{connectors}.
\textit{Transformation nodes} define the operations performed on input data, and the output(s) generated from them.
Each transformation node may accept multiple data streams as input, and may generate multiple data streams as output.
Moreover, they propagate metadata from input data streams to output data streams, and append metadata about the transformation itself to preserve provenance.
\textit{Visualization nodes}, on the other hand, may accept multiple data streams as input, but instead of generating output streams, they generate visualizations.
In this work, we primarily use Vega \cite{satyanarayan2015reactive} to declare visualization nodes in JSON format.
Additionally, a node itself can be defined as a workflow, allowing users to form hierarchies of workflows.
This allows users to re-use existing workflows to form more complex workflows, which also serves as a form of abstraction 
(see Figure~\ref{fig:subflow} for examples).

\begin{figure}[b]
\centering
\fbox{
\includegraphics[width=.95\linewidth]{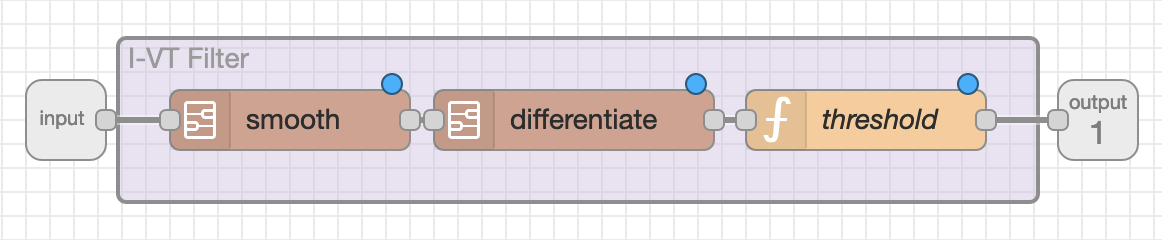}
}
\caption{
\textbf{IVT sub-flow on Node-RED:}
\textnormal{
Incoming gaze data first enters \textit{Smooth}, which removes high-frequency noise via a Butterworth filter.
Next, it enters \textit{Differentiate}, which calculates first derivatives via a Savitzky-Golay filter.
Next, it enters \textit{Threshold}, which outputs fixation/saccade-labeled data using derivatives.}
\label{fig:ema-subflow}
}
\end{figure}

\subsection{Operations Dashboard}

The operations dashboard allows users to monitor active workflows, generate interactive visualizations, and perform data-stream control actions (see Figure~\ref{fig:ema-dashboard} for a sample dashboard).
We use Node-RED to implement the operations dashboard.
When designing a workflow, users may add visualization nodes at any desired point in the workflow.
Each visualization node, in turn, would generate a dynamic, reactive visualization on the operations dashboard.
In terms of data-stream control, we propose to include five data-stream control actions: start, stop, pause, resume, and seek.
These actions would enable users to navigate data streams temporally and perform visual analytics \cite{keim2008visual}, a particularly useful technique to analyze high-frequency, high-dimensional data.

\subsection{Platform Heuristics}

As a part of this study, we propose two heuristics to quantify the workload in streaming workflows:
1)~\textit{fluidity}, a ratio of inbound to outbound data frequency, and 2)~\textit{growth factor}, a ratio of inbound to outbound data volume.
These measures are similar in spirit to buffer underrun/overrun probability and packet jitter in streaming applications, with more contextual relevance to SWFs.\\

\noindent\textbf{Fluidity (F)}:
We propose Fluidity as a heuristic for computational bottlenecks in a transformation \mbox{$T:S_{i}{\rightarrow}S_{o}$}, where $S_{i}$ is the set of input streams, and $S_{o}$ is the set of output streams.
Here, each $s_o \in S_{o}$ will have an expected frequency $f_e$, and a mean observed frequency $f_\mu$.
Formally, we define Fluidity as,
\[
F(s_o,t)=1-\sqrt{1-({f_\mu}/{f_e})^2}
\]
where, $F(s_o,t)\in[0,1]$, and $f_\mu\in[0,f_e]$.
For any output stream $s_o$, $f_e$ is constant, but $f_\mu$ varies with time.
As $f_\mu$ decreases, the fluidity drops.
For larger drops in $f_\mu$, the drop in fluidity is higher.
The value of $f_\mu$ depends on factors including, but not limited to, hardware, concurrency, parallelism, and scheduling.
Intuitively, $f_\mu$ can be estimated by computing the number of samples generated by $T$ in unit time.
However, since this estimate can be noisy, a Kalman filter could be used.
Overall, a fluidity $\ll 1$ is an indicator of computational bottlenecks, while a fluidity $\approx 1$ is an indicator of good performance.
Bottlenecked transformations identified in this manner can be improved by either code-level optimization or executing in a faster runtime, which, in turn, would increase fluidity.\\

\noindent\textbf{Growth Factor (GF)}:
We propose Growth Factor as a heuristic for the change in data volume through a transformation \mbox{$T:S_{i}{\rightarrow}S_{o}$}, where $S_{i}$ is the set of input streams, and $S_{o}$ is the set of output streams.
Formally, we define Growth Factor as,
\[
GF(s_i,s_o)=\big(\sum\limits_{s{\in}S_o}{V(s)}\big)\big/\big(\sum\limits_{s{\in}S_i}{V(s)}\big)~,~V(s)=f_s\sum\limits_{c_i}{w_{c_i}}
\]
where, $GF(s_i,s_o)\in[0,\infty)$, and $V(s)\geq0$.
For any stream $s$, the data volume $V(s)$ is calculated using its frequency $f_s$, and the ``word size'' $w_{c_i}$ of each channel $c_i$ in $s$.
Here, the word size $w_{c_i}$ represents the number of bits occupied by a sample of data from channel $c_i$ (e.g., if $c_i$ is of type \textit{int32}, then $w_{c_i}=32$).
Furthermore, $GF<1$ indicates a data compression, $GF>1$ indicates a data expansion, and $GF=1$ indicates no change.
Intuitively, outputs from transformations with $GF<1$ are likely candidates for caching and transmitting over networks, as they output a lower volume of data than they receive.

%% file: tables/dds.attrs.tex
\begin{tabular}{ll}
\toprule
\mc{2}{c}{\textbf{DATA SOURCE}}                                                        \\
\midrule
\textbf{info}       & version, timestamp, and checksum (for identification)         \\
\textbf{device}     & model, manufacturer, and category of data source              \\
\textbf{fields}     & dtype, name, and description of all fields                    \\
\textbf{streams}    & information on all data streams generated from it             \\
\midrule
\mc{2}{c}{\textbf{DATA ANALYTIC}}                                                      \\
\midrule
\textbf{info}       & version, timestamp, and checksum (for identification)         \\
\textbf{sources}    & pointer(s) to the data source metadata                        \\
\textbf{fields}     & pointer(s) to the fields used in analysis                     \\
\textbf{inputs}     & pointer(s) to the streams used in analysis                    \\
\textbf{streams}    & information about all streams generated via analysis          \\
\midrule
\mc{2}{c}{\textbf{DATA SET}}                                                           \\
\midrule
\textbf{info}       & version, timestamp, and checksum  (for identification)        \\
\textbf{name}       & name of the data set                                          \\
\textbf{description}& description of the data set                                   \\
\textbf{keywords}   & keywords describing the data set (for indexing)               \\
\textbf{authors}    & name, affiliation, and email of each data set author          \\
\textbf{sources}    & pointer(s) to the data source metadata                        \\
\textbf{fields}     & pointer(s) to the fields used in analysis                     \\
\textbf{groups}     & viewpoints to query different slices of the dataset           \\
\textbf{resolver}   & path to an executable that resolves data in the dataset       \\
\bottomrule
\end{tabular}

%% file: sections/4.experiments.tex
Using StreamingHub, we build two interactive stream analysis workflows in the domains of \textit{eye movement analysis} \cite{jayawardena2020pilot,duchowski2020low} and \textit{weather analysis} \cite{dzodom2020keeping}. 
Here, different domains were used to test if StreamingHub generalizes across applications serving different scientific purposes.
From this, we evaluate three aspects of StreamingHub.
\begin{itemize}
\item Can we use DDS metadata to replay stored datasets?
\item Can we build workflows for domain-specific analysis tasks?
\item Can we create domain-specific data visualizations?
\end{itemize}

Using three datasets, we evaluate StreamingHub on two distinct stream analysis tasks in the domains of \textit{eye movement analysis} and \textit{weather analysis}.
Here, we describe our data source(s) using the proposed metadata format, and build stream processing workflows that leverage this metadata.
Based on our observations, we discuss the utility of StreamingHub for data stream processing, and uncover challenges that inspire future work.

\begin{figure*}[t]
\centering
\fbox{
\includegraphics[width=.95\linewidth]{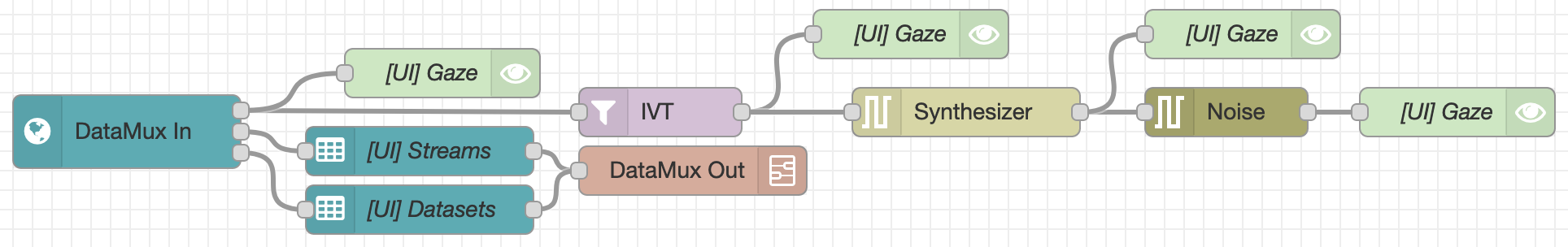}
}
\caption{
\textbf{Eye movement analysis workflow on Node-RED:}
\textnormal{Gaze data from the \textit{DataMux In} node enters an \textit{IVT} node, which identifies fixations/saccades.
This data enters a \textit{Synthesizer} node, which synthesizes gaze data.
Next, this data enters a \textit{Noise} node, which adds Gaussian noise.
The four \textit{[UI]~Gaze} nodes visualize gaze data, fixation/saccade data, synthetic gaze data, and noisy gaze data.}
\label{fig:ema-workflow}}
\end{figure*}

\subsection{Eye Movement Analysis}

\subsubsection{Data Preparation}
We use two existing datasets, \textit{ADHD-SIN} \cite{jayawardena2020pilot} and \textit{N-BACK} \cite{duchowski2020low}, each providing gaze and pupillary measures of subjects during continuous performance tasks.
We first pre-processed these datasets to provide normalized gaze positions (x,y) and pupil diameter (d) over time (t).
Any missing values were filled via linear interpolation, backward-fill, and forward-fill, in order.

\subsubsection{Experiment Design}
In this experiment, we perform three tasks:
1)~\textit{replay} eye movement data,
2)~obtain eye movement \textit{analytics} in real-time, and
3)~observe data/analytics through eye movement \textit{visualizations}.

\begin{figure*}[ht]
\centering
\fbox{
\includegraphics[width=.95\linewidth]{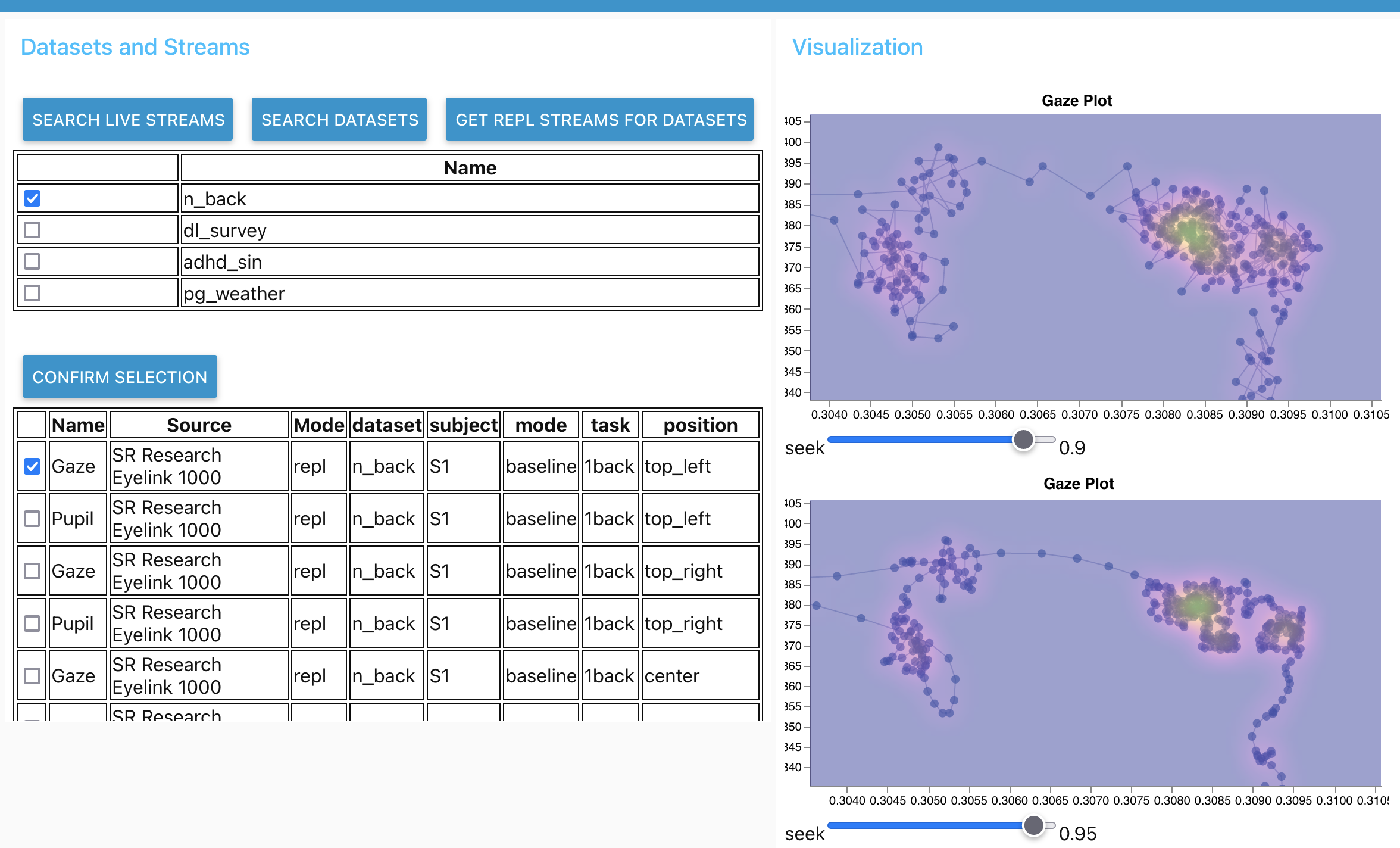}
}
\caption{
\textbf{Operations Dashboard for Eye Movement Analysis:}
\textnormal{It visualizes, in real-time, raw (top-right) and synthesized (bottom-right) gaze data from replayed data (bottom-left) of the N-BACK dataset (top-left).
The top-left buttons list available data-streams and datasets.
The \textit{"Confirm Selection"} button (mid-left) subscribes to selected data-streams and visualizes incoming data.}
\label{fig:ema-dashboard}
}
\end{figure*}

\vspace{1em}
\noindent\textbf{Task 1: Replay:}
We first generate DDS metadata for the N-BACK and ADHD-SIN datasets.
Then we implement a \textit{resolver} (using Python) for each dataset, which maps queries into respective data files, and references them in the metadata.
Next, we use the DataMux API to list the data-streams of each dataset, and subscribe to them.

\vspace{1em}
\noindent\textbf{Task 2: Analytics}:
We create an eye movement analysis workflow on Node-RED, using visual programming.
We begin by creating empty sub-flows for each transformation, and wiring them together, as shown in Figure~\ref{fig:ema-workflow}.
Next, we implement these sub-flows to form the complete workflow.
For instance, we implement the I-VT algorithm \cite{salvucci2000identifying} in the IVT sub-flow (see Figure~\ref{fig:ema-subflow}) to classify data points as fixations or saccades.

\vspace{1em}
\noindent\textbf{Task 3: Visualization}:
We implement an interactive 2D gaze plot using the Vega JSON specification \cite{satyanarayan2015reactive}.
It visualizes gaze points as a scatter plot, connects consecutive gaze points using lines, and overlays a heat map to highlight the distribution of gaze points across the 2D space.
It also provides a \textit{seek} bar to explore gaze data at different points in time.
The resulting operations dashboard, as shown in Figure ~\ref{fig:ema-dashboard}, displays the datasets and data-streams available for selection, the selected data-streams, and a real-time visualization of their data.

\begin{figure*}[t]
\centering
\fbox{
\includegraphics[width=.95\linewidth]{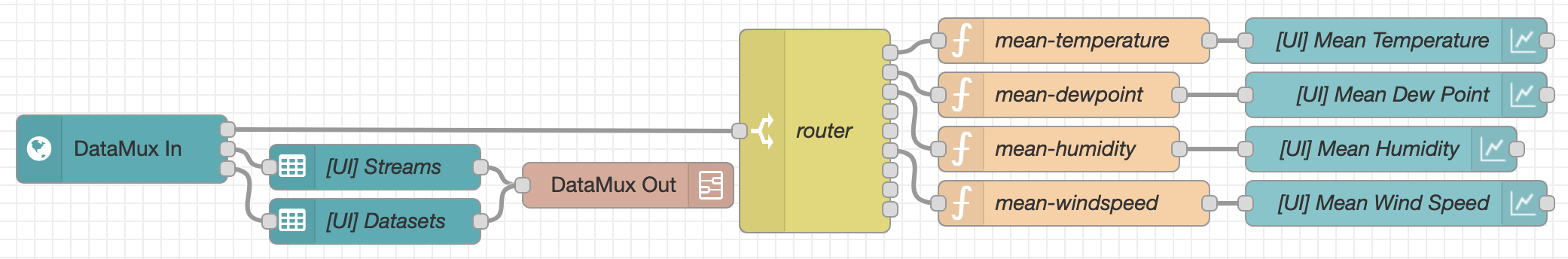}
}
\caption{
\textbf{Weather analysis workflow on Node-RED:}
\textnormal{Weather data from the \textit{DataMux In} node is passed into a \textit{Router} node, which routes data (by type) into relevant visualization nodes.
Only \textit{temperature}, \textit{dew point}, \textit{humidity}, and \textit{wind speed} data types are visualized.}
\label{fig:wp-workflow}
}
\end{figure*}

\subsection{Weather Analysis}

\subsubsection{Data Preparation}
We use a dataset from PredictionGames \cite{dzodom2020keeping} providing daily min/mean/max weather statistics (e.g., temperature, dew point, humidity, wind speed) of 49 US cities between 1950 and 2013.
We first pre-processed this dataset by splitting weather data into seperate files by their city, and ordering records by their date.

\subsubsection{Experiment Design}
In this experiment, we perform three tasks:
1)~\textit{replay} weather data from different cities,
2)~calculate moving average \textit{analytics} in real-time, and
3)~observe data/analytics through chart-based \textit{visualizations}.

\begin{figure*}[ht]
\centering
\fbox{
\includegraphics[width=.95\linewidth]{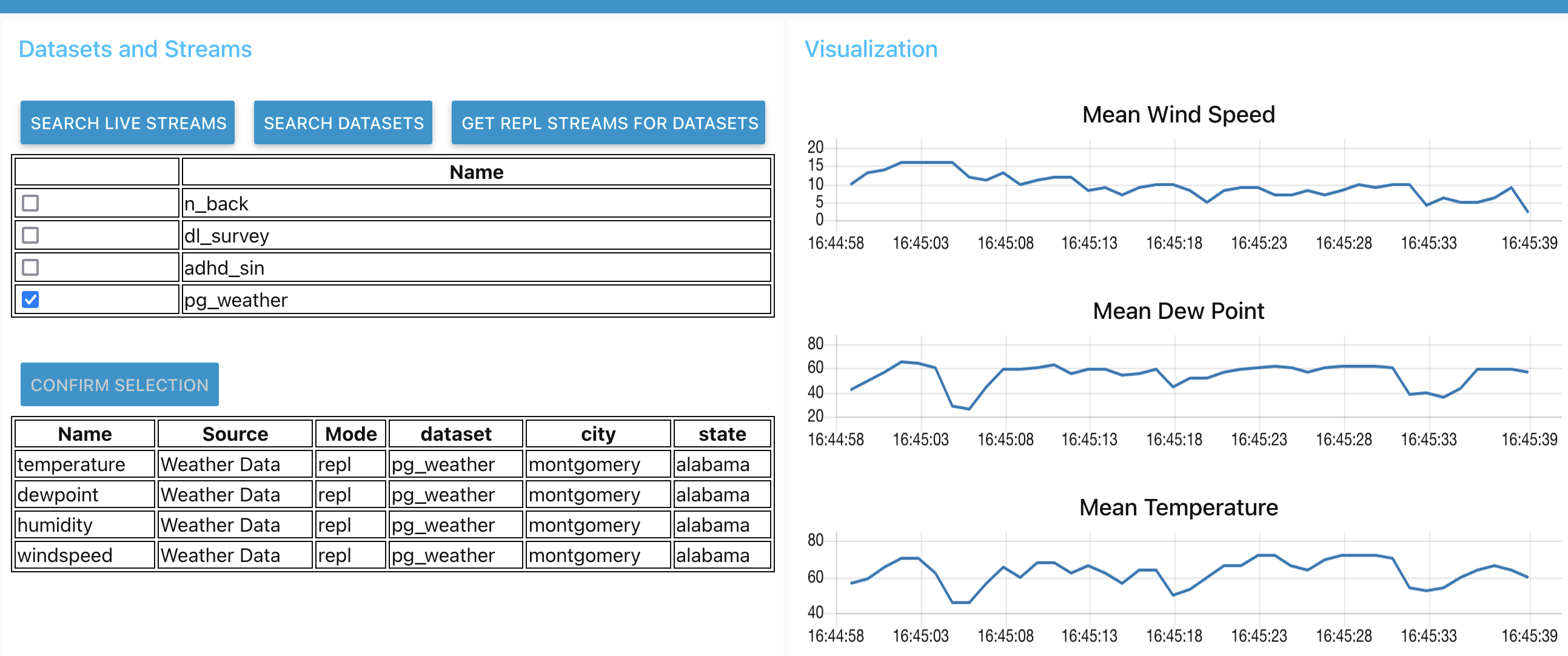}
}
\caption{
\textbf{Operations Dashboard for Weather Analysis:}
\textnormal{It visualizes, in real-time, the mean values of \textit{wind speed}, \textit{dew point}, and \textit{temperature} (right).
All buttons and selection controls (left) offer the same functionality as described in Figure~\protect\ref{fig:ema-dashboard}.}
\label{fig:wp-dashboard}
}
\end{figure*}

\vspace{1em}
\noindent\textbf{Task 1: Replay:}
We first create DDS metadata for the weather dataset.
Here, we set the replay frequency as 1 Hz to speed up analysis (i.e., 1 day = 1 second).
Then we implement a \textit{resolver} (using Python) for this dataset, which maps queries into respective data files, and references them in the metadata.
Next, we use the DataMux API to list the data-streams of the dataset, and subscribe to them.

\vspace{1em}
\noindent\textbf{Task 2: Analytics:}
We create a workflow to perform moving average smoothing on weather data and visualize each metric as shown in Figure~\ref{fig:wp-workflow}.
For the scope of this experiment, we only use the temperature, dew point, humidity, and wind speed metrics.

\vspace{1em}
\noindent\textbf{Task 3: Visualization:}
Here, we accumulate and visualize the averaged weather data through interactive charts.
Figure~\ref{fig:wp-dashboard} shows a snapshot of the operations dashboard while visualizing replayed weather data.
It shows the datasets and data-streams available for selection, the selected data-streams, and a real-time visualization of their data.

%% file: sections/5.evaluation.tex


\begin{table*}[t]
\centering
\caption{Mean latency ($\bar{t}$), inbound data size ($V_i$), and outbound data size ($V_o$) of four transformations ($T$) ranked by increasing compute demand ($D_c$) and outbound data size ($D_s$), in both \textit{REPLAY} and \textit{LIVE} modes.\label{tab:performance}}
\input{tables/performance.tex}
\end{table*}

\vspace{1em}
In Table~\ref{tab:performance}, we quantitatively assessed the performance of different sections of the workflows.
We picked four transformations (i.e., \textit{nodes}) from the workflows of both experiments, and ranked them in increasing order of their compute demand and outbound data size.
Next, we sent 100,000 samples into each node at 50\,Hz, and measured the mean latency $\bar{t}$, total inbound data size $V_i$, and total outbound data size $V_o$.
We obtained these measurements in both \textit{replay} and \textit{live} modes (see Table~\ref{tab:performance}).

Here, $\bar{t}$ was higher for all transformations in \textit{live} mode, compared to \textit{replay} mode.
This may be due to differences in the underlying implementation; in \textit{live} mode, data is proxied from devices using LabStreamingLayer whereas in \textit{replay} mode, data is read (and cached) from files, and the frequency at which samples are replayed is regulated by a timer.
This setup may be prone to I/O overhead, as seen in the results.
Moreover, in both modes, the inbound and outbound data sizes are large since they are serialized in \textit{JSON} format.
In the future, we plan to use a space-efficient serialization format such as \textit{ProtoBuf} to address this limitation.

\section{Heuristic Evaluation}
Here, we compare the proposed heuristics with the metrics reported in Table~\ref{tab:performance}.
For each transformation in this table, we send $100,000$ samples at $50\,Hz$, calculate the proposed heuristics, and compare our heuristics with the reported metrics.
Based on our results, we determine if the proposed heuristics are consistent with the reported metrics.

\begin{table}[ht!]
\centering
\caption{Comparison of performance metrics ($\bar{t}$, $V_i$, $V_o$) with heuristics ($F$, $GF$) for each transformation ($T$) on Table~\ref{tab:performance}, in REPLAY mode.\label{tab:heuristics}}
\input{tables/heuristics.tex}
\end{table}

Table~\ref{tab:heuristics} shows the mean latency $\bar{t}$ (ms), fluidity $F$, inbound data size $V_i$, outbound data size $V_o$, and growth factor $GF$ values obtained.
For transformations with $F<1$, $\bar{t}$ was higher compared to transformations with $F=1$.
Moreover, transformations with $V_o~\ll~V_i$ had $GF<1$; transformations with $V_o~\gg~V_i$ had $GF>1$; and transformations with a data-dependent $V_o$ had no $GF$ value.

Fluidity is based on observed frequency, and is not impacted by latency until throughput is affected.
Furthermore, fluidity varies with time, but growth factor is independent of time.
Thus, growth factor can be pre-calculated and used to dynamically optimize workflows for constrained resources.
Moreover, both heuristics can be applied at transformation-level to determine which outputs to cache or re-generate.
In this work, we only evaluated the proposed heuristics in replay mode, and not in live mode.
A comprehensive evaluation should thus be performed to validate their utility across different applications.

%% file: tables/performance.tex
\begin{tabular}{l|cc|l|ccc|ccc}
\toprule
\mr{2}{$T$}                               &\mr{2}{$D_c$} &\mr{2}{$D_s$} &\mr{2}{$*$}                     &\mc{3}{c|}{\textbf{REPLAY MODE}}   &\mc{3}{|c}{\textbf{LIVE MODE}}     \\
                                          &              &              &                                &$\bar{t}$ (ms) &$V_i$   &$V_o$     &$\bar{t}$ (ms) &$V_i$   &$V_o$     \\
\midrule
\textbf{Mean} (Weather A)                 &1             &1             & w,s=50                         &0.018          &782\,MB &15.64\,MB &0.024          &782\,MB &15.64\,MB \\
\textbf{Threshold} (Eye Movement A)       &2             &2             & I-VT filter (v=80)             &0.097          &782\,MB &438\,MB &0.132            &782\,MB &322.9\,MB \\
\textbf{Derivative} (Eye Movement A)      &3             &4             & {int32}$\rightarrow${float64}  &0.243          &782\,MB &1.54\,GB  &0.391          &782\,MB &1.54\,GB  \\
\textbf{Smooth} (Eye Movement A)          &4             &3             & S-G filter (w=7)               &0.645          &782\,MB &781.9\,MB &0.726          &782\,MB &781.9\,MB \\
\bottomrule 
\end{tabular}

%% file: tables/heuristics.tex
\begin{tabular}{l|cc|ccc}
\toprule
$T$                  &$\bar{t}$ (ms) &$F$    &$V_i$      &$V_o$      &$GF$   \\
\midrule
\textbf{Mean}        &0.018          &1.000  &782\,MB    &15.64\,MB  &0.02   \\
\textbf{Threshold}   &0.097          &0.999  &782\,MB    &438\,MB    & --    \\
\textbf{Derivative}  &0.243          &0.934  &782\,MB    &1.54\,GB   &2.00   \\
\textbf{Smooth}      &0.645          &0.742  &782\,MB    &781.9\,MB  &1.00   \\
\bottomrule
\end{tabular}

%% file: sections/6.discussion.tex
\subsection{Societal Impact}
One foreseeable application of this framework is to test experimental setups, both pre and post-collection.
For pre-collection, one could simply stream in either random data or simulate test cases to verify that workflows run as expected.
For post-collection, one could replay data through a workflow and verify that experimental results match.
Overall, this provides an ecosystem to build robust, error-tolerant, and most importantly, reproducible real-time applications.

\subsection{Limitations}

In this study, our goal was to conceptualize \textit{metadata propagation} and demonstrate how it facilitates reusable, reproducible analytics in scientific workflows with zero supervision.
For this reason, our evaluation was focused on functionality and performance heuristics, and not user experience.
In the future, we plan to conduct a user study to obtain feedback on (a) the workflow design experience and (b) the benefits of added reusability and reproducibility.

When developing our framework, we relied on the DDS metadata schema, which was purpose-built for metadata propagation.
While creating DDS metadata for live data is relatively straightforward, doing so for stored data requires additional effort.
In particular, it requires one to code resolver scripts that map queries to data.
Despite being a one-time effort, this may limit the adoption of our framework outside real-time settings.

\subsection{Future Work}

\noindent\textbf{Promoting Reuse:}
In this study, we implemented domain-specific sub-flows to execute our case studies (e.g., Fixation Detector, Gaze Synthesizer).
In the future, we plan to refine these sub-flows and make them publicly available.
Through this, we aim to promote the reuse of domain-specific workflows.

\noindent\textbf{Improving Communication}:
In our framework, we used LabStreamingLayer to create, discover, and subscribe to data streams.
In the future, we plan to explore alternative communication methods including, but not limited to, message brokers like \textit{MQTT} \cite{hunkeler2008mqtt}, and messaging protocols like \textit{protobuf} \cite{blyth2019proio}.

\noindent\textbf{Time Synchronization}:
Our framework currently relies on LabStreamingLayer to synchronize different data streams in time.
Internally, LabStreamingLayer uses the Precision Time Protocol (PTP) to perform this.
In the future, we plan to implement such protocols as Node-RED sub-flows to promote flexibility and reuse.

\begin{figure}[ht!]
\centering
\fbox{\includegraphics[width=.97\linewidth]{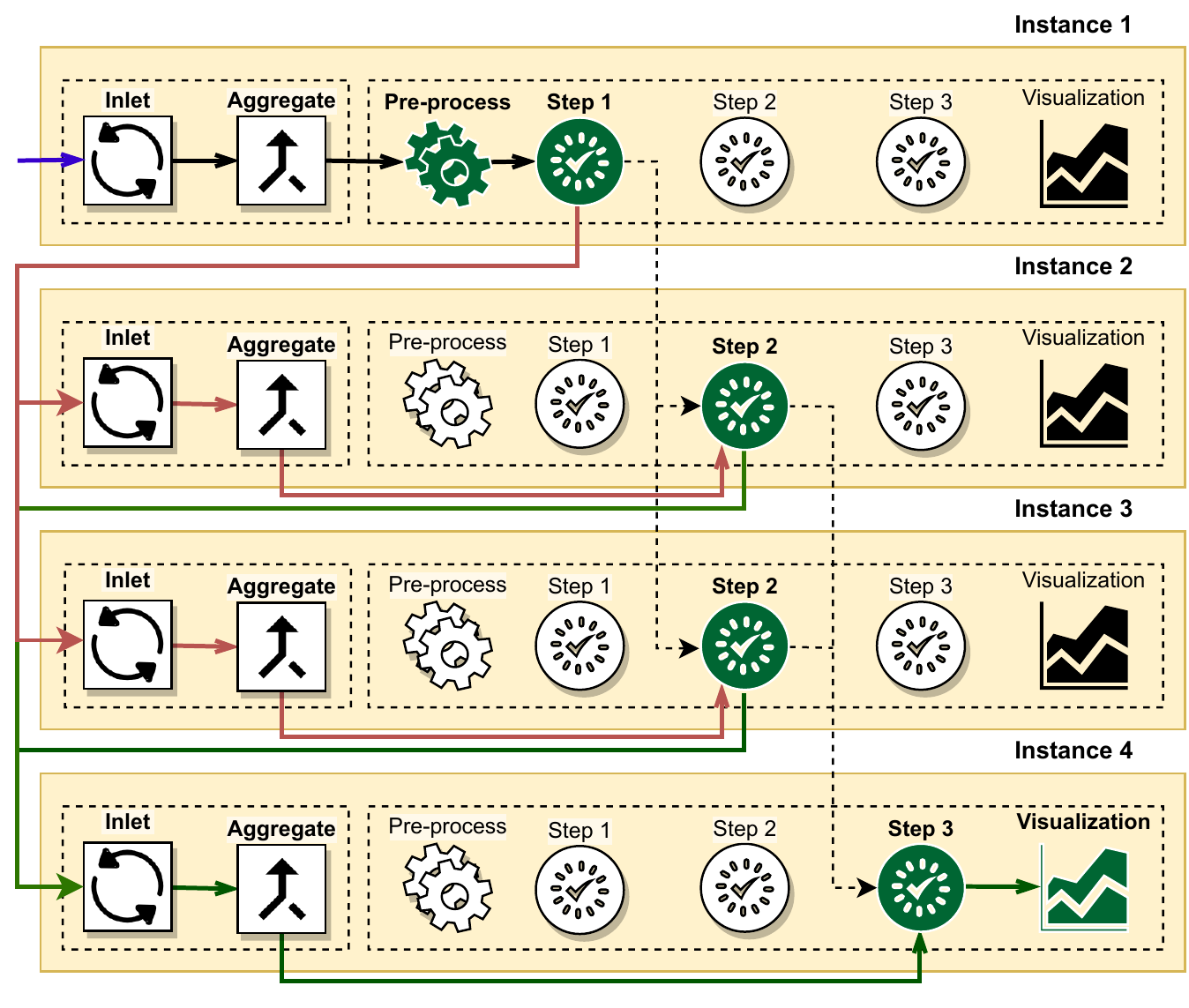}}
\caption{
Distributed Execution of Workflows
\textnormal{with \textit{workflow~chaining} (solid arrows) and \textit{workload~sharing} (dotted arrows)}.
\label{fig:streaminghub-distributed}}
\end{figure}

\noindent\textbf{Distributed Execution:}
At present, our framework only supports standalone workflow execution.
In the future, we plan to integrate a workflow engine (e.g., \textit{Apache Flink}, \textit{Amazon Kinesis}) to enable distributed execution, and evaluate two aspects of it:
\textit{workflow chaining} (where analytic outputs of workflows are consumed by secondary workflows), and
\textit{workload sharing} (where components of workflows are executed across nodes) (see Figure \ref{fig:streaminghub-distributed}).
Moreover, since workload distribution is better optimized with knowledge of the underlying data, workload, and compute power \cite{isah2019survey}, we plan to leverage \textit{fluidity} and \textit{growth factor} heuristics for such optimization.

%% file: sections/7.conclusion.tex
We developed a framework (StreamingHub) to build stream analysis workflows that intelligently consume data, propagate metadata to downstream tasks, and thereby auto-generate reusable, reproducible analytic outputs with zero supervision.
We also developed a metadata format (DDS) which facilitates metadata propagation in this framework, and proposed two heuristics to quantify computational aspects of a workflow built using it.
We discussed how we implemented this framework using DDS, Node-RED, LabStreamingLayer, and WebSockets.
We explained how it propagates metadata, and how it facilitates replaying datasets as data streams, passing data streams into workflows, performing data stream control, and conducting exploratory data analysis.
We applied this framework for two case studies: \textit{eye movement analysis}, and \textit{weather analysis}.
For eye movement analysis, we developed a workflow to detect fixations, simulate gaze data from fixations, and visualize gaze data.
For weather analysis, we developed a workflow to visualize weather-related statistics, and replay data at different frequencies than recorded.
We showed that the proposed heuristics indicate computational bottlenecks and estimate the expansion/compression of data in workflows.
To promote reuse, our code is publicly available on \url{https://github.com/nirdslab/streaminghub}.